\documentclass[twocolumn,showpacs,preprintnumbers,amsmath,amssymb]{revtex4}

\usepackage{graphicx}
\usepackage{dcolumn}
\usepackage{bm}
\usepackage{booktabs}

\begin{document}

\preprint{APS/123-QED}

\title{Pattern formation in oscillatory complex networks
consisting of excitable nodes}

\author{Xuhong Liao$^1$, Qinzhi Xia$^1$, Yu Qian$^2$, Lisheng Zhang$^1$}
\author{Gang Hu$^1$}
\email{ganghu@bnu.edu.cn}
\author{Yuanyuan Mi$^1$}
\email{miyuanyuan0102@163.com}
\affiliation{$^{1}$ Department of
Physics, Beijing Normal University, Beijing 100875, China;\\
$^{2}$ Nonlinear Research Institute, Baoji University of Arts and
Sciences, Baoji 721007, China}%

\date{\today}

\begin{abstract}
Oscillatory dynamics of complex networks has recently attracted
great attention. In this paper we study pattern formation in
oscillatory complex networks consisting of excitable nodes. We find
that there exist a few center nodes and small skeletons for most
oscillations. Complicated and seemingly random oscillatory patterns
can be viewed as well-organized target waves propagating from center
nodes along the shortest paths, and the shortest loops passing through both
the center nodes and their driver nodes play the role of oscillation
sources. Analyzing simple skeletons we are able to understand and
predict various essential properties of the oscillations and
effectively modulate the oscillations. These methods and
results will give insights into pattern formation in complex
networks, and provide suggestive ideas for studying and controlling oscillations in
neural networks.
\end{abstract}

\maketitle

\section{Introduction}
Many social and natural systems have been well described by complex
networks\cite{Watts1998,Barabasi1999,Barabasi2009,Souza2009}.
Complex networks with excitable local dynamics have
attracted particularly great attention for their wide applications,
such as epidemic spreads\cite{Kuperman2001,Pastor2001}, chemical
reactions\cite{Tinsley2005}, biological
tissues\cite{Sporns2004,Muller2008}, among which neural networks are
typical examples\cite{Sporns2004,Muller2008,Roxin2004,Sinha2007}.
Complexity of network structures and excitability of local
dynamics are two major characteristics of neural
networks\cite{Gerstner2002,Izhikevich2004}.
Oscillations in these networks determine rich and important physiological
functions\cite{Gray1994,Buzaski2004}, such as visual
perception\cite{Usrey1999}, olfaction\cite{Stopfer1997}, cognitive
processes\cite{Ward2003}, sleep and arousal\cite{Steriade1993}.
Therefore, oscillations in neural networks and other excitable networks
have been studied extensively.

Problems of pattern formation in these excitable systems call for further investigation,
because early works on pattern
formation focused on patterns in regular
media\cite{Meron1992,Cross1993,Pearson1993}. It is natural to ask
what pattern formation looks like in complex networks, and whether
there are some common rules in
different types of networks. It's very recently, Turing patterns in
large random networks have been discussed by Hiroya Nakao \emph{et
al.}\cite{Nakao2010}. In the present paper we study
another type of pattern, self-sustained oscillatory patterns in
complex networks consisting of excitable nodes, which are
important in physics, chemistry and biology. Since
each excitable node cannot oscillate
individually\cite{Fitzhugh1961}, there must exist some delicate
structures supporting the self-sustained
oscillations\cite{Roxin2004,Sinha2007,Lewis2000,Steele2006}. So far,
some concepts, such as recurrent
excitation\cite{Donovan1999,Sanchez2000,Roxin2004,Lewis2000},
central pattern
generators\cite{Selverston1985,Yuste2005,Vogels2005}, have been
proposed to describe these structures. However, if networks consist of
large numbers of nodes and random interactions, it's difficult to detect these
structures\cite{Sinha2007,Steele2006}. In previous
papers\cite{Liao2011,Qian2010}, we proposed a method of dominant phase
advanced driving (DPAD) to study the phase relationship between
different nodes based on oscillatory data. Oscillation sources for
self-sustained oscillations are identified successfully.
However, the topological effects on the dynamics are not fully understood.

The interplay between the topological connectivity and the network
dynamics has become one of the central topics under
investigation\cite{Strogatz2001,Albert2002,Boccaletti2006}. The
present paper is to explore the mechanism of pattern formation in
oscillatory excitable networks and unveil the topological dependence
of the oscillations. This paper is organized as follows. Section II
introduces the excitable networks of B\"{a}r Model. Simulation
results are provided in section III, where center nodes and target
waves are identified. In section IV, the skeletons of different
oscillations are displayed to unveil the topological effects on
network dynamics. In section V results in previous sections are extended to
networks with different sizes and degrees. Section
VI gives extensions to excitable scale-free networks. Networks with
Fitzhugh-Nagumo Model as local dynamics are also discussed.
The conclusions are given in the last section VII.

\section{Model of Networks}
We consider complex networks consisting of $N$ excitable nodes. The
network dynamics is described as follows:

\begin{eqnarray}
\frac{du_{i}}{dt} &=&-\frac{1}{\varepsilon }u_{i}(u_{i}-1)(u_{i}-\frac{%
v_{i}+b}{a})+D_{u}w_i,  \notag \\
\frac{dv_{i}}{dt} &=&f(u_{i})-v_{i}, \quad i=1,2,\ldots , N  \notag\\
w_{i}
&=&\frac{\sum_{j=1}^{N}M_{ij}u_{j}}{\sum_{j=1}^{N}M_{ij}u_{j}+K}-u_i,\\    
\quad f(u_{i})&=&\left\{
\begin{array}{cl}
0, & u_{i}< \frac{1}{3}, \\
1-6.75u_{i}(u_{i}-1)^{2}, & \frac{1}{3}\leq u_{i}\leq 1, \\
1, & u_{i}>1.%
\end{array}%
\right.  \notag
\end{eqnarray}%
B\"{a}r model is adopted as local dynamics\cite{Bar1993}, where
parameters $\{a, b, \varepsilon\}$ are properly set so that each
node possesses excitable local dynamics. The adjacency matrix
$M_{ij}$ is defined by $M_{ij}=1$ if node $i$ is connected with node
$j$ and $M_{ij}=0$ otherwise. Coupling $w_{i}$ represents the total
interaction on a given node $i$ from all its neighbor nodes.
This form of coupling is used to ensure that any excited node can
excite its rest neighbor nodes with proper
values of $D_u$ and $K$. Other
forms of coupling, which have similar effects, are also feasible,
such as diffusive coupling. This type of interaction has been widely
used in neural models\cite{Muller2008,Roxin2004,Lewis2000} and other
excitable networks\cite{Greenberg1978,Kuperman2001,Bak1990}. During the
simulations, different types of networks are generated and
the connections between different nodes are bidirectional and
symmetric. For simplicity, we study at first
homogeneous random networks with an identical degree $k$, i.e.\ each
node interacts with an equal number of $k$ nodes randomly chosen.
Meanwhile, we assume that all nodes have identical
parameters so that any heterogeneity in network patterns is not due
to the topological inhomogeneity, but results from the
self-organization in nonlinear dynamics. In the present paper we
focus on the self-sustained periodic oscillations.

\section{Center nodes and Target Waves}
The homogeneous random network studied is displayed in Fig.\
1(a). With the parameters given, the system has a large probability
(about $95\%$) to approach periodic oscillations from random initial
conditions. Moreover, different initial conditions approach
different oscillations in most cases. For instance, we observed $961$
different oscillations within $1,000$ tests, and the other samples
reached the rest state. The evolution of average signals
$<u(t)>=1/N\sum_{i=1}^{N}{u_{i}(t)}$ for three different
oscillations A, B, C is displayed in Fig.\ 1(b). These three oscillations have
different periods ($T_A=5.99$, $T_B=8.37$, $T_C=7.00$
with $T_{A,B,C}$ being periods of oscillations A, B, C,
respectively). In Figs.\ 1(c), (d) and (e) spatial snapshots of
oscillations A, B and C are plotted, respectively. All these
patterns have seemingly random phase distributions, in which
the structures supporting the oscillations are deeply hidden.

\begin{figure*}[b]
\centering
\includegraphics[scale=1.0]{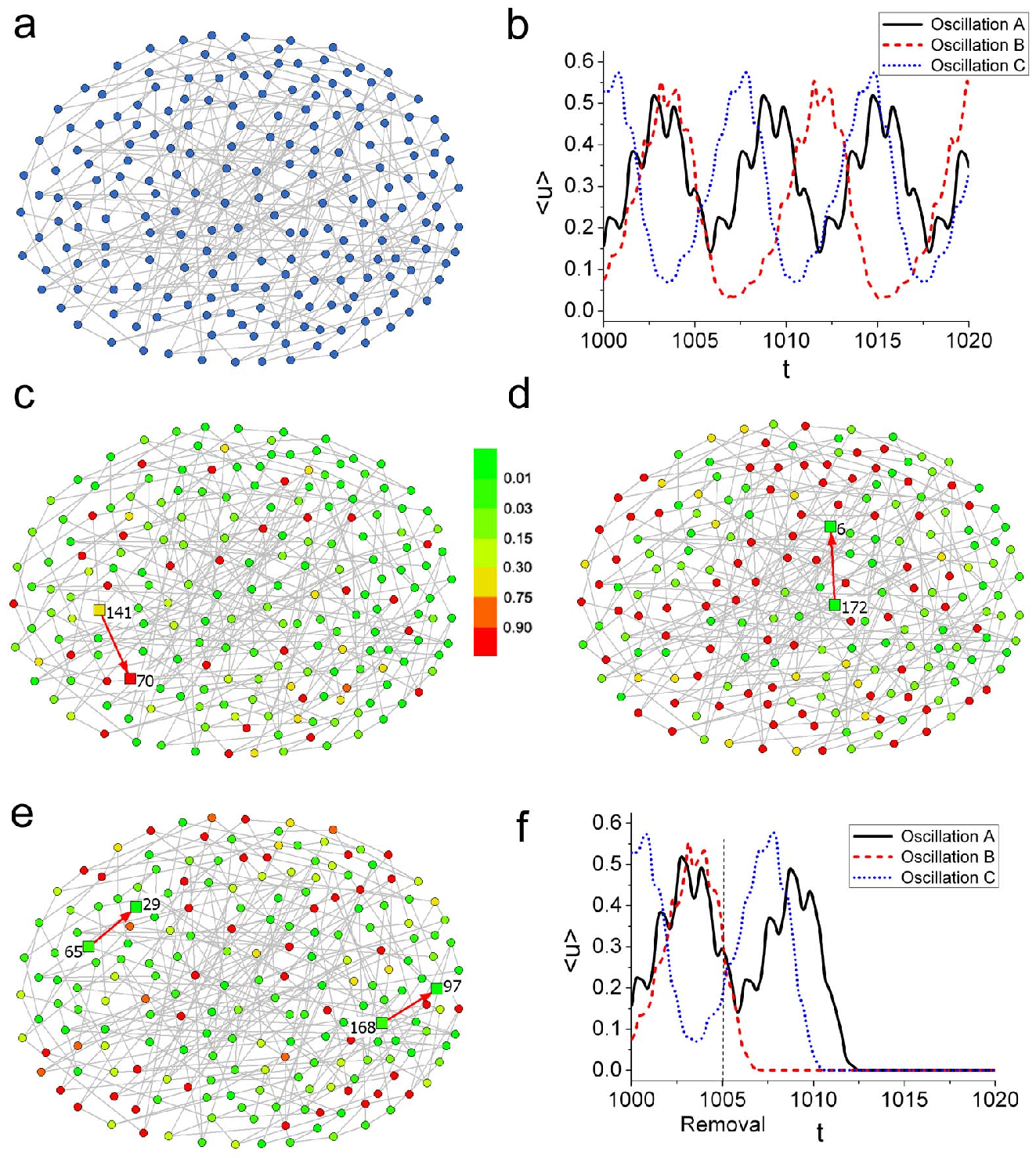}

\caption{\label{Fig1} (Color online) (a) Homogeneous random network
studied. $N=200$, $k=3$. The network dynamics is described by
equation (1) with parameters $a=0.84$,
$b=0.07$, $\varepsilon =0.04$, $D_u=0.4$, $K=0.8$. All these
parameters are used in Figs.\ 1-5. (b) Periodic time evolution of $<u(t)>=%
1/N \sum_{i=1}^N u_i(t)$ for three different random initial
conditions. These three oscillations are denoted as
oscillations A, B and C, respectively. (c) Snapshot of oscillation
A at a certain time. The phase distribution among
different nodes seems random. In the following figures, all
snapshots are displayed with local variable $u$ plotted without
further remarks and the arrowed links in all figures
represent the dynamical driving relationship. (d)(e) Snapshots of
oscillations B and C, respectively. (f) Suppression of
oscillations A, B and C by removing square nodes $70$, $6$, and
($29$, $65$) at $t=1005$, respectively. The time of node removal is
denoted by the vertical dash line.}

\end{figure*}

We start our analysis from local dynamics of excitable networks.
Because each node is an excitable system\cite{Fitzhugh1961,Bar1993},
the individual node will stay at the rest state forever without
perturbation. Since there is no external pacemaker in the network,
there must be some loops to support the self-sustained oscillations,
where nodes can be repeatedly excited in sequence. Therefore, it
is natural to conclude that the topological loop structure of
complex networks is crucial for the network oscillations\cite{Lewis2000,Steele2006,Liao2011,Qian2010}. However,
in complex networks there are extremely large numbers of loop sets
(for the network in Fig.\ 1(a) with $N=200$ nodes and $M=300$
interactions, there are $2^{101}-1$ loop sets\cite{Bollobas1998}). A
crucial question is which loop set plays the essential role for a
given oscillation. Because nodes in the network are excited in sequence,
 all waves propagate forward along the shortest
paths\cite{Muller2008,Lewis2000,Steele2006}. The loops dominating
the oscillations must obey this `` shortest path" rule, which means
the source loops should be as short as possible. Furthermore,
due to the existence of the refractory period, these loops must also be
sufficiently large to maintain the recurrent excitation. Here, the
problem remained is how to reveal these shortest loops.

We study the above loop problem by making perturbation to each oscillation and observe the system's
response. A few nodes randomly chosen are removed from the network
at each test. (Here removing a node means discarding all
interactions of this node.) In most cases the oscillation is robust.
However, we find in surprise that the oscillation is crucially
sensitive to some specific nodes. These specific nodes for a given
oscillation are defined as key nodes, among which a minimum number of nodes
can be removed to suppress the oscillation. In Figs.\
1(c), (d) and (e) different key nodes for oscillations A, B and C
are displayed with large squares, respectively. Both oscillations A and B can
be suppressed by just removing one key node, as shown in Fig.\
1(f). However, we can never suppress oscillation C by removing any
single node. There are two pairs of key nodes displayed in Fig.\ 1(e).
In order to terminate oscillation C (see also Fig.\
1(f)), we have to remove two key nodes simultaneously, one from the pair ($29,65$) and the other
from the pair ($97,168$). The diverse behavior displayed in
Figs.\ 1(c), (d) and (e) indicates that even though the parameter distributions and the node
degrees are homogeneous in the network, the dynamical patterns have
delicate and heterogeneous self-organized structures where
different nodes play significantly different roles in the
oscillations.

We find further that all key nodes for these oscillations appear in
directly interacted pairs. In each pair one node drives the other,
i.e. $141 \rightarrow 70$, $172 \to 6$, and $65 \to
29$, $168 \to 97$. (The bidirectional link between nodes $i$
and $j$ is denoted by an arrowed link $i\to j$, if the
interaction from node $i$ is favorable for exciting node $j$
from the rest state.) Considering the crucial influence of key nodes
on the oscillations, we suggest that the function of the driven
nodes is to excite the whole network, while the function of the
driving ones is to keep their partners oscillating. Thus
these driven nodes ($70$ for A, $6$ for B, $29$ and $97$ for C) are regarded as
center nodes for the oscillations while the driving ones are regarded as the
drivers of the center nodes. An oscillation with $n$ centers is
called $n$-center oscillation. Both oscillations A and B are
one-center oscillations, while oscillation C is a two-center oscillation.

\begin{table}
\caption{\label{Table1} Number of center nodes for different
oscillations in Homogeneous Random  Networks (HRNs).
Parameters ($a,b,\varepsilon,D_u$) are set the same as Fig.\ 1,
 except constant $K$ ($K=0.8$ for HRNs with $N=100, 200$ and
 $K=1.8$ for other networks). One thousand different networks are investigated with random initial
conditions for the statistics in each column.}
\begin{ruledtabular}
\begin{tabular}{ccccccc}
N & 100 & 200 & 200 & 1,000 & 2,500 & SF200\footnotemark[1]\\
k & 3   & 3   &5 &  7    &  8    & \textless k\textgreater=4\\
\hline
Periodic oscillations & 785 & 955 & 675 & 549 & 328 & 172\\
one-center &290 & 98 & 191 & 457 & 298 & 156\\
two-center &244 & 172& 170 & 85  & 30  & 15\\
three-center &144 &199& 97  & 7   & 0   & 1
\\
four-center &52  & 189& 73  & 0   & 0   & 0 \\
others\footnotemark[2]   &55  &297& 144 & 0   & 0   & 0\\
\end{tabular}
\end{ruledtabular}
\footnotemark[1]{Scale-free networks with average degree $<k>=4$.}
\footnotemark[2]{Periodic oscillations with more than four centers.}
 \end{table}

The existence of key nodes and center nodes is general for
periodic oscillations in excitable complex networks.
We investigated Eq.\ (1) with random initial conditions for different networks and sampled stable periodic oscillations. The transient time for each oscillation depends on the network size $N$. When the network size increases, the transient will be prolonged.
Moreover, the transient time is also effected by the type of the
pattern. Generally speaking, the more center nodes the pattern has,
 the longer the transient needs to be. When the oscillation reached stability, center nodes were identified. Numbers of center nodes for most oscillations are listed in Table I. For other oscillations remained, we did not make a further
search, because identifying more than four center nodes is very computationally consuming. Anyway, we find that most oscillations have self-organized structures with an extremely small number of center nodes. Thus the features of oscillations A, B, and C can be identified as the typical behavior of self-sustained oscillations in excitable complex
networks.

Because of the significant effects of center nodes on oscillations,
we expected that the source loops of the oscillations must be around the
center nodes. Further study confirmed the
expectation. We identified that there are just some well-organized loop
structures around the center nodes to maintain the self-sustained
oscillations. Two
principles are proposed for pattern formation in a given network oscillation.

(i)  Waves propagate forward from center nodes to the whole network
along the shortest paths.

(ii) The shortest loops passing through both center nodes and their drivers
play the role of oscillation sources and dominate the oscillation
behavior.

With these two principles we can clearly
reveal oscillation sources, illustrate wave propagation paths and
unveil the topological effects on the oscillations.

\begin{figure*}
\centering
\includegraphics[scale=1.0]{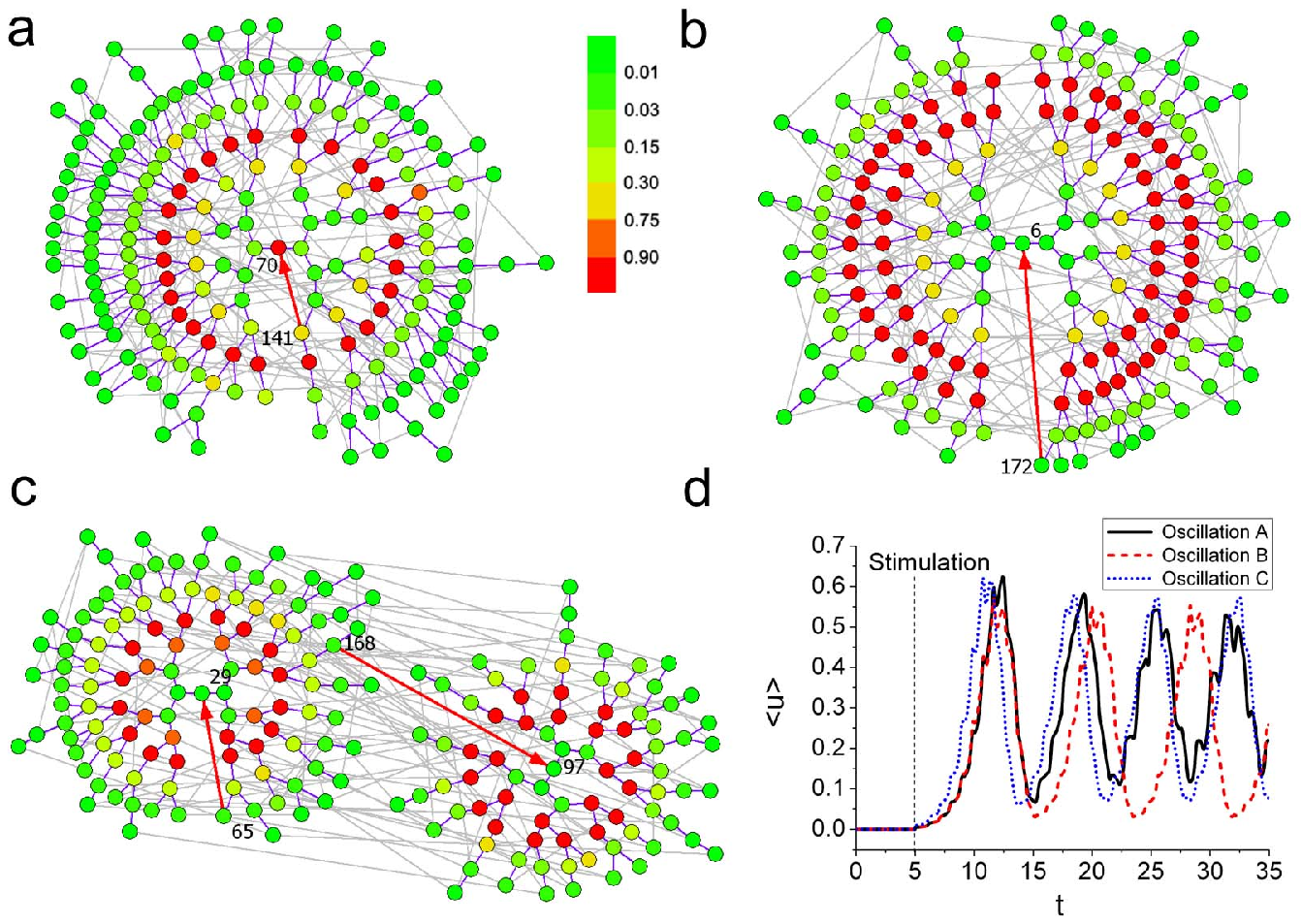}

\caption{\label{Fig2}  (Color online) (a) Snapshot same as Fig.\ 1(c)
for oscillation A, with each node placed
around center node $70$ according to the distance from the center. The
purple (dark) lines show the shortest paths from center node $70$ to
the other nodes, while the grey (light) lines mean all the other
interactions between different nodes. Perfect target
waves are observed propagating from center node $70$. The
bold arrow denotes the driving from node $141$ to center
node $70$. (b) Snapshot same as Fig.\ 1(d) for oscillation B.
The target center is node $6$ while its driver is node
$172$. (c) Snapshot same as Fig.\ 1(e) for oscillation C, with nodes placed
around two centers $29$ and $97$. Two-center target pattern is
identified. (d) Creation of oscillations A, B,
C by initially stimulating few center nodes (node $70$ for A, node
$6$ for B, nodes ($29$,$65$) for C). Average signals $<u(t)>$ for
different oscillations are displayed. Stimulations are performed
at $t=5$ denoted by the vertical dash line.}
\end{figure*}

Based on the first principle, we can demonstrate the oscillatory
pattern for each oscillation according to a simple placing rule as follows.  At
first, place each center node at a certain position. Second, if there is only one center node, locate all the other nodes around this center according to the
distances (shortest paths) from it. However, if there are two
synchronous centers, two clusters of nodes will exist, each around a center.
The other nodes should select the cluster with
the ``nearest" center node before the rearrangement. During the cluster selection if a
node has the same distances from both centers, it can be
included to either cluster. This simple placing rule transforms
all random patterns into well-behaved target waves. Similar operation can be applied
to oscillations with more centers. Snapshots of oscillation A, B, and C in new
order are displayed in Figs.\ 2(a), (b)
and (c), respectively, which are exactly the same as those in Figs.\
1(c), (d) and (e). In these figures surprisingly
well-ordered target waves are observed, one-center target waves for
oscillations A and B, and two-center target waves for oscillation C,
which are in sharp contrast with the random phase distributions in
Figs.\ 1(c), (d) and (e). All nodes are driven by waves emitting from center nodes and the importance of the center nodes are demonstrated clearly. The recurrent excitation of the center nodes via the driving key nodes is the reason why the center nodes can
keep oscillating to excite the whole network. It is instructive to
observe these self-sustained target waves in oscillatory random
networks and demonstrate how these waves self-organize. On the basis of these target patterns, different oscillation sensitivities observed in Fig.\ 1 can be understood. First, due to
the one-center target structure of Figs.\ 2(a) and 2(b), we can
definitely terminate the oscillation by removing a center
node ($70$ for A, $6$ for B), since the center node is the only wave
source. The oscillation can also be suppressed by removing the
driving node ($141$ for A, $172$ for B) because the driving node
is the only driver of the center node, without which the center can
no longer oscillate sustainedly. Second, since oscillation C has a
structure of two-center target waves, removing any single key node
can not destroy the oscillation sources completely. Both target
centers (or their drivers) should be removed simultaneously to
suppress this two-center oscillation. Similarly, in order to terminate an
oscillation with $n$ centers, $n$ centers (drivers) should be
removed simultaneously.

In Fig.\ 1(f) effective suppression of given
oscillations is displayed. However, how to create a given oscillation with high efficiency is still not clear.
Excitable networks, such as that in Fig.\ 1(a), have a huge number of
attractors, each of which has a small basin of attraction. If we try
to reach a given oscillation by random initial conditions we may
need thousands or even millions of tests which are
computationally consuming and practically unreasonable. However,
when the center nodes and their drivers are identified we
can recover a given oscillation with high efficiency by manipulating
only a very few nodes. To create oscillation A (B) from the all-rest state we
only need to initially stimulate single center node $70$ ($6$) while
the interaction from the center node to its driving key node $141$
($172$) is blocked during the initial excitation period of the center nodes.
We find that the excitation activities propagate away from the center
node, and then come back via the driving key node to reexcite the
center node. Then the system evolves autonomously to target
pattern A (B) via the self-organized excitation propagation in the
network. Generally speaking, we can recover any given $n$-center target
pattern by initially stimulating $n$ centers with the interactions
from these centers to their drivers blocked during the initial
excitation periods of center nodes. In the following paper, this
excitation procedure is briefly called $n$-center node excitation,
without additional remarks on the interaction modulations. In
Fig.\ 2(d) we present the evolution generated by
one-node-excitation with the solid (dash) curve, which recovers
oscillation A (B) asymptotically. In order to recover oscillation
C, both center nodes $29$ and $97$ should be excited simultaneously.
The creation of oscillation C is shown by the dotted curve in Fig.\
2(d).

\section{Skeletons and Oscillation Control}
Based on Principle (ii), we can construct a skeleton and reveal the oscillation
source for each oscillation by analyzing the network topology.
The skeleton of a given oscillation
means a subnetwork consisting of some short topological loops passing through both
the center nodes and their drivers. Topological effects on a network
oscillation can be well unveiled based on the skeleton. In Figs.\
3(a), (b) and (c) skeletons of oscillations A, B and C are
displayed, respectively. In Fig.\ 3(a) we display all topological
loops with length $L \leqslant 10$, passing through the pair of key nodes
($70$, $141$). In Fig.\ 3(b) the skeleton of oscillation B is
plotted, consisting of loops with length $L \leqslant 11$ passing through
the key node pair ($6$, $172$). An interesting difference
between Figs.\ 3(a) and 3(b) is that the shortest loop in Fig.\ 3(a) ($%
L_{min}(A)=5$) is much smaller than that in Fig.\ 3(b)
($L_{min}(B)=9$). In Fig.\ 3(c) we show the skeleton of oscillation
C consisting of loops with length $L \leqslant 10$, passing through the pair
of nodes ($29$, $65$) or ($97$, $168$). The skeleton supporting oscillation C consists of two clusters with $L_{min}(C)=7$. Furthermore, for each oscillation under investigation the shortest loops displayed in the skeleton always have
successive driving relationship. In Figs.\ 3(a)-(c), these
successive driving shortest loops are indicated by arrows. These
driving loops, supporting self-sustained oscillations of the center
nodes, are regarded as the oscillation generators. The phenomenon in Fig.\ 1(b) that
oscillations A, B and C have different periods ($T_A=5.99$,
$T_B=8.37$, $T_C=7.00$) can be understood
from these oscillation generators. It has been known that a pulse
can circulate along a one-dimensional (1D) loop consisting of
excitable nodes. The period of the oscillation increases as
the loop's length increases\cite{Winfree1991}. Since the shortest
loop in each skeleton dominates the oscillation, we have the conclusion
$T_B>T_C>T_A$ for $L_{min}(B)>L_{min}(C)>L_{min}(A)$. That's the reason
why different oscillations may have different periods.

\begin{figure*}
\centering
\includegraphics[scale=1.0]{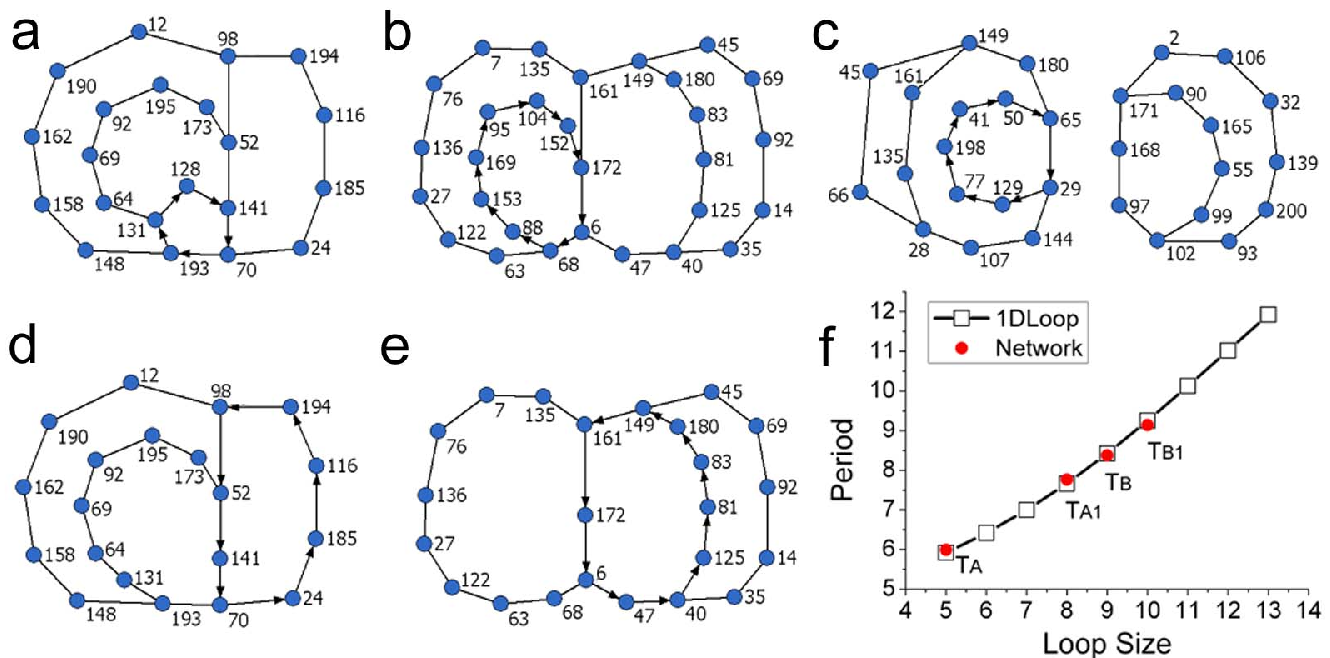}

\caption{\label{Fig3}  (Color online) (a)(b)(c) Skeletons of
oscillations A, B and C, respectively. (a) Skeleton of oscillation A
consisting of all loops passing through key nodes ($70$, $141$) with length
$L \leqslant 10$. (b) Skeleton of
oscillation $B$ consisting of all loops passing through key nodes ($6$, $172$) with length $L \leqslant 11$%
. (c) Skeleton of oscillation $C$ consisting of all loops passing through the key node pair (%
$29$, $65$) or the other pair ($123$, $140$) with $L \leqslant 10$. (d)
Skeleton of oscillation $A_1$ with node $128$ removed from
oscillation A. (e) Skeleton of oscillation $B_1$ with node $88$
removed from oscillation $B$. (f) Oscillation periods versus the
driving shortest loops identified. The white squares in the
solid curve represent numerical results for $1D$ loops, which have the same
parameters as those in Fig.\ 1. $T_A$, $T_B$, $T_{A_1}$, $T_{B_1}$
(red (dark) circles) denote periods of oscillations $A$, $B$ and
their modulated oscillations $A_1$, $B_1$. All circles are located
around the squares.}

\end{figure*}

The structures of skeletons in Figs.\ 3(a), (b) and (c) are greatly
simplified in contrast with the original complex network in Fig.\ 1(a). They
contain much less number of nodes, and reduce the original
high-dimensional complex structures to various sets of 1D loops. It
is importance to find these small skeletons which indicate many
essential features of the network oscillations. We
can efficiently modulate the oscillations just by analyzing these
simple skeletons. In the following discussions the oscillation period is taken as a measurable quantity to demonstrate the oscillation modulations.

 At first we modify oscillation A by removing node $%
128$. This operation changes oscillation A to oscillation $A_{1}$.
Based on Fig.\ 3(a) we can predict the network evolution after the modulation.
First, although the shortest $5$-node loop is
destroyed, there are still some other loops containing
center $70$ and its driver $141$. The oscillation will be maintained. Second, the new shortest loop among the
remaining loops will emerge as a dynamical loop, which guarantees the
recurrent excitation of center $70$ and maintains the network
oscillation. Because the length of the new shortest loop ($70 \to
24\to 185 \to 116 \to 194 \to 98
\to 52 \to 141 \to 70$) is $8$, we expect
that the modified oscillation $A_{1}$ must have a larger period.
Our predictions are confirmed. The
skeleton of oscillation $A_{1}$ is shown in Fig.\ 3(d) where the right loop
(marked by the arrowed loop) actually emerges as the oscillation
generator. And the period of oscillation $A_1$ is indeed larger than that
of oscillation A ($T_{A_{1}} = 7.77>
T_A = 5.99$). Similar operations are applied to oscillation B.
Oscillation $B_1$ is obtained by removing single node $88$ from
oscillation B. Analyzing the skeleton in Fig.\ 3(b) we expect that
this operation must prolong the original period $T_B$ to $T_{B_1}$
($T_{B_1}>T_B$), for the new shortest loop has a length $L=10$. In
Fig.\ 3(e) the skeleton of oscillation $B_1$ is displayed as
expected. Then we find $T_{B_1} = 9.14> T_B = 8.37$. In Fig.\ 3(f)
periods of 1D oscillatory loops with
different sizes are displayed with white squares in the solid curve. Periods of network
oscillations A, B, $A_1$ and $B_1$ are also displayed with red (dark) circles. Both sets of
periods coincide well. It demonstrates that
simplified skeletons indicate essential features of
complicated patterns, and the shortest loops passing through
both the center nodes and their drivers indeed dominate the dynamics
of the network oscillations.

\begin{figure*}
\centering
\includegraphics[scale=1.0]{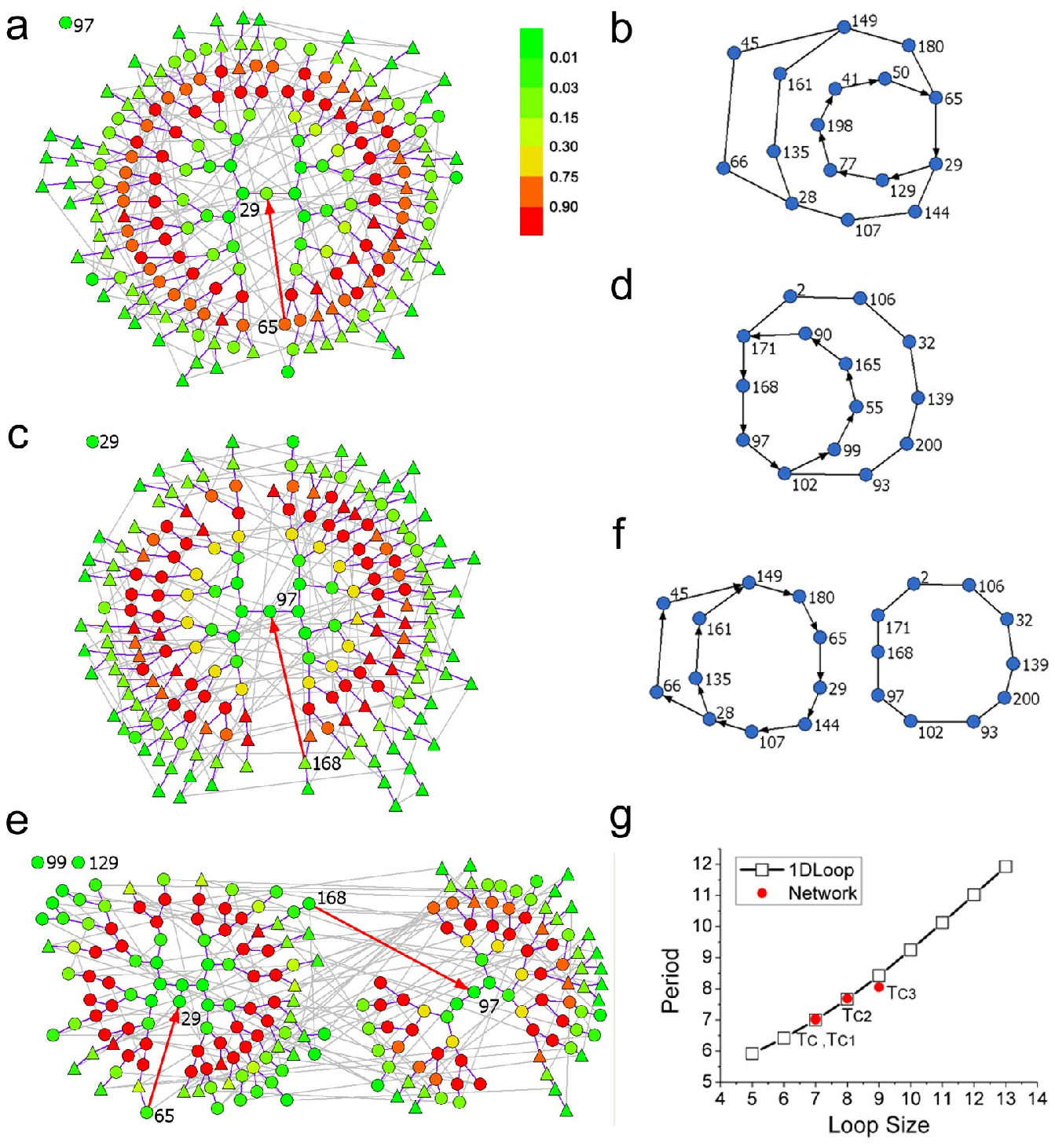}

\caption{\label{Fig4} (Color online) (a) Snapshot of oscillation $C_1$
with center node $97$ removed from oscillation $C$ (see Fig.\ 2(c)).
Triangles denote the nodes migrating between different clusters after
the modulation. (b) Skeleton of oscillation $C_1$. (c) Snapshot of
oscillation $C_2$ with center node $29$ removed from oscillation
$C$. (d) Skeleton of oscillation $C_2$. (e) Snapshot of oscillation
$C_3$ with two nodes ($129$, $99$) removed from oscillation C
simultaneously. (f) Skeleton of oscillation $C_3$. (g) Periods $T_C$ and $T_{C_i}$, $i=1, 2, 3$ (red (dark)
circles) versus the lengths of the shortest loops in
the skeletons. All data of $T_{C_i}$ can be approximately
predicted by the results of 1D loops (white squares in the solid curve).}

\end{figure*}

The modulation diversity can be much richer for oscillations with
more centers. Different modulations are applied to oscillation $C$. In the subsequent
paragraphs, responses of oscillation C to the removal of different nodes,
(i) node $97$, (ii) node $29$, (iii) nodes $129$ and
$99$, will be studied. We find that all simulations of the
network oscillations fully coincide with predictions from the simple
skeleton in Fig.\ 3(c).

(i) If center node $97$ is removed, the network oscillation must
change, i.e. from oscillation C to $C_1$. Analyzing the skeleton of
oscillation C, the right sub-skeleton must be destroyed by removing
its center $97$ while the left sub-skeleton is remained intact to
support oscillation $C_1$. Since only left target center $29$ works,
the original two-center target pattern must be transformed to a
one-center target pattern, and the nodes in the original right
cluster must move to the left cluster. The left cluster will grow
from the boundary with nodes migrating from the destroyed
cluster. We present the target pattern of oscillation $C_1$ in Fig.\
4(a) by simulation, and find a pattern the same as we predicted. In
Fig.\ 4(b) we plot the skeleton of oscillation $C_1$ which is
nothing but the left sub-skeleton in Fig.\ 3(c). (ii) If center node
$29$ is removed from oscillation C, the left sub-skeleton in Fig.\
3(c) is destroyed. The resulting oscillation is denoted by $C_2$.
We expect that node $97$ will work as the only center and the shortest loop in
right sub-skeleton will work as the oscillation generator. In Figs.\
4(c) and 4(d) we observe that all predictions are fully confirmed.
(iii) If two nodes are removed simultaneously, node $129$ from the
left cluster and node $99$ from the right one, oscillation $C_3$ is
generated. We predict from Fig.\ 3(c) that the two-center target
pattern should be maintained (since the functions of two centers are
preserved) and the skeleton of $C_3$ can be deduced from Fig.\
3(c) with the shortest loops in both clusters destroyed. Then we
stimulate oscillation $C_3$ and plot a snapshot by
arranging the nodes in order. Two-center target waves
are verified in Fig.\ 4(e). The skeleton of oscillation $C_3$ is
displayed in Fig.\ 4(f). Both Figs.\ 4(e) and 4(f) fully confirm the
above predictions.

Meanwhile, the above operations have adjusted the oscillation
periods. In case (i) since the oscillation generator of $C_1$
remains the same as that of oscillation C, the resulting period
$T_{C_1}$ should remain approximately the same as $T_C$. For
oscillation $C_2$ in case (ii), the new oscillation generator (the
arrowed shortest loop) in Fig.\ 4(d) has length $8$, and then period
$T_{C_2}$ should increase to about $7.66$ by comparing with 1D loop
data in Fig.\ 4(g). Since loop nodes $129$ and $99$ are removed,
oscillation $C_3$ has the shortest source loops of length $9$
(arrowed loops in the left cluster in Fig.\ 4(f)). Thus period
$T_{C_3}$ should be close to $8.42$ (see Fig.\ 4(g)), which is
considerably larger than $T_C$. In Fig.\ 4(g)
numerical results of the modulated networks are compared with those of 1D loops.
Both sets of data agree well with each other. It is amazing that by
removing node $97$ we dramatically change the oscillation pattern
while keeping the period almost unchanged. In contrast, by removing
two nodes ($129$, $99$) we keep the two-center target pattern while
largely slowing down the oscillation. All these seemingly strange
responses can be well explained with the skeleton in Fig.\
3(c).

So far, we discussed oscillations A, B, C in the given network Fig.\
1(a) in detail. However, oscillation patterns in a complex network
are much more abundant. The choice of key nodes and related source
loops depends on initial conditions, because the basins of
attraction of different attractors may be very complicated in
nonlinear dynamic systems. For a homogeneous random network all
nodes are topologically equivalent and each node may play a role of
a center node or the driver of the center node. The only condition
is that the shortest loops passing through both the center node and the driver
must be large enough to guarantee the recurrent excitation.

\section{Complex networks with different sizes and
degrees}

Till now we focused on Eq.\ (1) with $N=200$ and $k=3$. All
characteristics observed in this particular case can be extended to
networks with different sizes and degrees. Here we
study another example of Eq.\ (1) with $N=400$ and $k=4$. The
network structure is displayed in Fig.\ 5(a). With a certain initial
condition we observe an oscillation $\Phi$ with a snapshot shown in
Fig.\ 5(b). This oscillatory pattern has $7$ key nodes, which are
displayed with squares in Fig.\ 5(b). Four centers ($4$, $57$,
$176$, $260$) are identified. Removing four key nodes simultaneously
from four different sets, i.e.\ one node from each set, we can suppress
this oscillation. This process is displayed by the solid curve in
Fig.\ 5(c). Different from oscillations A, B and C, in Fig.\ 5(b) only three sets
of key nodes appear in pairs (such as ($57$, $339$),
($176$, $13$), ($260$, $382$)), while key node $4$ appears without
any partner. The reason is following.
Since each node has a degree $k=4$, a center node may
have a single dynamical driver (such as $339 \to 57$, $13
\to 176$, $382 \to 260$), or multiple drivers (such as
node $4$ in Fig.\
5(d), having two drivers $244$ and $360$). If a center node has only one driver, the driver node also
becomes a key node for controlling the center node.
However, when the center node has multiple drivers, removing one of
these drivers can not terminate the function of the center. Thus this
center node does not have a partner node for the oscillation
suppression. Similar to Fig.\ 2(d) we can generate oscillatory
pattern in Fig.\ 5(b) from the all-rest state by initially stimulating the
four centers ($4$, $57$, $176$, $260$) with interactions from these
centers to their drivers blocked during the initial excitation periods
of the center nodes. Time evolution of this oscillation generation is shown
in Fig.\ 5(c) with the dotted curve.

\begin{figure*}
\centering

\includegraphics[scale=1.0]{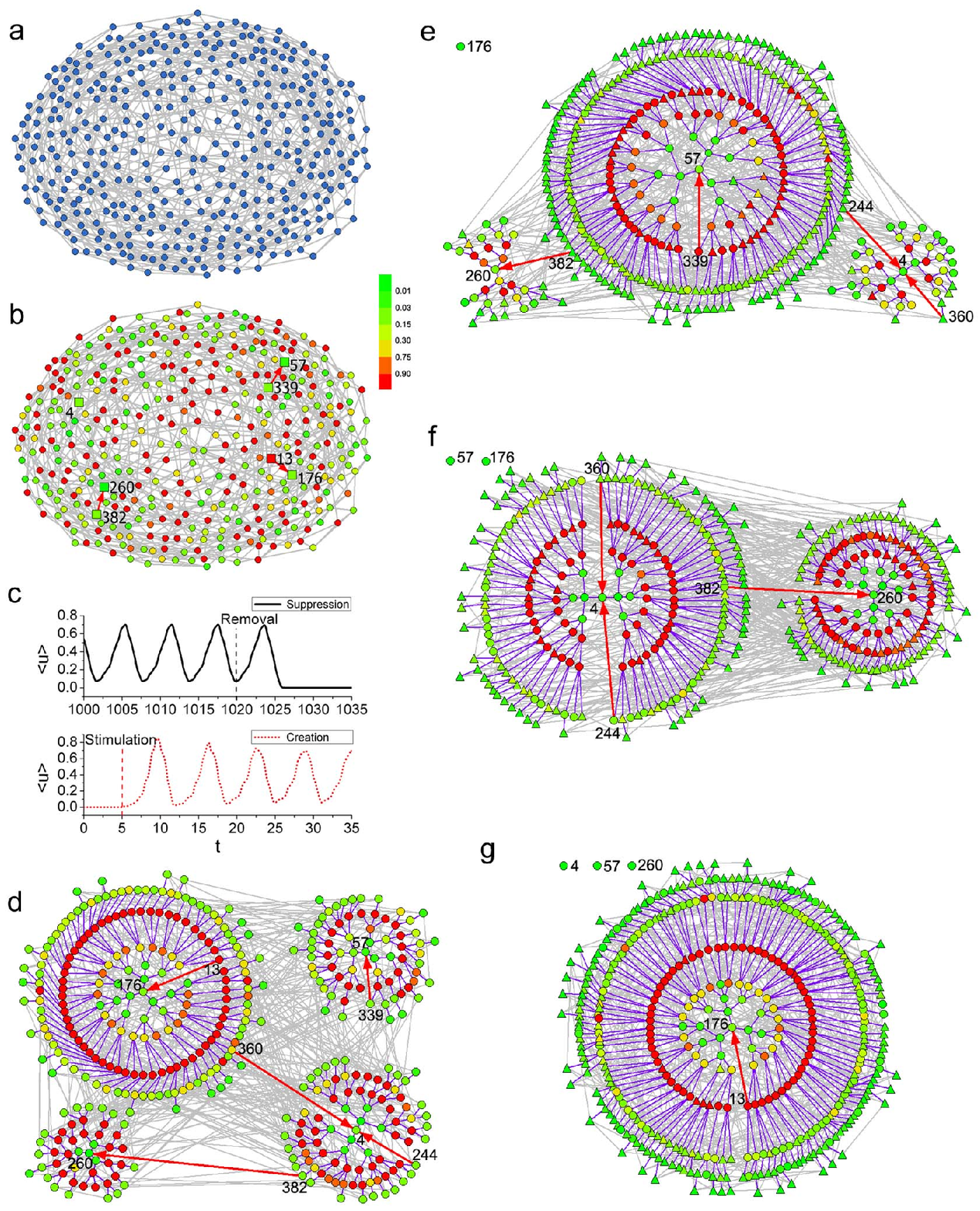}

\caption{\label{Fig5}  (Color online) (a) Homogeneous network with $N=400$,
$k=4$. The other parameters are the same as those in Fig.\ 1. (b)
Snapshot of oscillation $\Phi$ generated randomly. Four sets of
key nodes are displayed with squares. (c) Suppression and creation
of oscillation $\Phi$. Removing simultaneously four center nodes
($4$, $57$, $176$, $260$) can suppress the oscillation as
shown by the solid curve. Oscillation
$\Phi$ can also be generated by initially stimulating four center nodes ($4$, $57$, $176$,
$260$), as shown by the dotted curve below. Different operation
times are denoted by the vertical dash lines. (d) Snapshot same as (b) for
oscillation $\Phi$, with nodes are
rearranged in order. (e)-(g) Snapshots of oscillations after
different modulations to oscillation $\Phi$. Triangles mean the
nodes migrating between different clusters after the modulations.
(e) Snapshot of three-center oscillation with center $176$ removed
from oscillation $\Phi$. (f) Snapshot of two-center oscillation with
two centers ($57$, $176$) removed from oscillation $\Phi$. (g)
Snapshot of one-center oscillation with three centers ($4$, $57$,
$260$) removed from oscillation $\Phi$.}
\end{figure*}

In Fig.\ 5(d) we show exactly the same snapshot as that in Fig.\
5(b) with all nodes rearranged in four clusters according to their
distances from different centers, i.e., each node chooses the cluster
with the ``nearest" center node and then it is placed in
the selected cluster according to the distance from the center.
Different sizes of four clusters result from the asynchronous
excitation of different centers. If an oscillatory pattern has
multiple centers, each center emits excitation waves and controls a
cluster of nodes. A node will belong to the $i$th cluster if the
excitation wave from the $i$th center reaches this node first in
comparison with the other centers. Therefore, if all centers
have synchronous excitation any given node is controlled by the
nearest center as we did in Fig.\ 2(c). If multiple centers are not
synchronous, i.e., they are excited at different times, the
measurement of the distance should be modified by counting the
excitation time differences of various centers. In case of
oscillation $\Phi$, four centers are excited at slightly different
times. Specifically, in each round node $176$ is excited first,
nodes $57$ and $4$ have a single-step delay (one-step here means
$T/n$, with $T$ being the oscillation period and $n$ being the
number of nodes in a single wave length), while node $260$ has a
two-step delay. Then the ``nearest" center means the center node
with the shortest distance among ($d_1$, $d_2+1$, $d_3+1$, $d_4+2$),
while ($d_1$, $d_2$, $d_3$, $d_4$) being the actual topological
distances from centers ($176$, $57$, $4$, $260$) to the given node.
This method of distance measurement is applied to all the patterns
where more than one centers exist. With this arrangement we find
that the seemly random phase distribution in Fig.\ 5(b) is actually
a well-behaved four-center target wave pattern. All the modulations
to oscillation C shown in Fig.\ 4 can be applied to
oscillation $\Phi$ in Fig.\ 5(b). For instance, by removing center
$176$ we can transform the original four-center target waves to
three-center waves with centers $4$, $57$ and $260$. All nodes
migrating between different clusters after the modulations are also
displayed by triangles in Fig.\ 5(e). In Figs.\ 5(f) and 5(g) we
removed two center nodes ($57$, $176$) and three center nodes ($4$,
$57$, $260$), respectively. Two-center and one-center target
patterns are found, where all the remaining centers emit target
waves. All these modulation results show the generality of two
principles.

\section{Extensions}

In the previous discussions we consider only homogeneous random
networks where all nodes have the same degree. Both Principles (i)
and (ii) can be extended to Erd\"{o}s-R\'{e}nyi (ER) networks and
scale-free (SF) networks which are inhomogeneous in topological
structures. Results in these networks are similar. It
has been known that functional networks of the human brain exhibit
scale-free properties\cite{Sporns2004,Eguiluz2005}. In Fig.\ 6(a) we
present an example of SF network with $N=200$, $<k>=4$. The size of
each node $i$ is proportional to the nature logarithm of its degree $k_i$. For
this network we perform $1,000$ tests from different random initial
conditions and find $142$ self-sustained periodic oscillations.
Among these oscillatory patterns we identify $128$ oscillations with
single center, $13$ oscillations with two centers and $1$ oscillation with
three centers. The statistics for different networks is also listed
in Table I. These results confirm that the existence of a small number of center
nodes is also popular in inhomogeneous networks. In Figs.\ 6(b) and
(c) we present two snapshots of different oscillations (one-center
oscillation $SF_A$ and two-center oscillation $SF_B$) from different
initial conditions. The phase distributions seem complicated and
random. However, some key nodes and center nodes for the
oscillations are also identified (one pair of key nodes for
oscillation $SF_A$ and two pairs for oscillation $SF_B$). The given
oscillations can be suppressed (Fig.\ 6(d)) and created (Fig.\ 6(e))
by simply modulating the center nodes. In Figs.\ 6(f) and (g) we
plot exactly the same snapshots as those in Figs.\ 6(b) and (c),
respectively. With the placing rule, the random phase distributions of oscillations $SF_A$
and $SF_B$ can be rearranged to well-behaved one-center target waves (Fig.\ 6(f))
and two-center target waves (Fig.\ 6(g)), respectively. The skeleton of
oscillation $SF_A$ is shown in Fig.\ 6(h), based on which we can
make oscillation modulations as we did in Fig.\ 4.

\begin{figure*}
\centering
\includegraphics[scale=1.0]{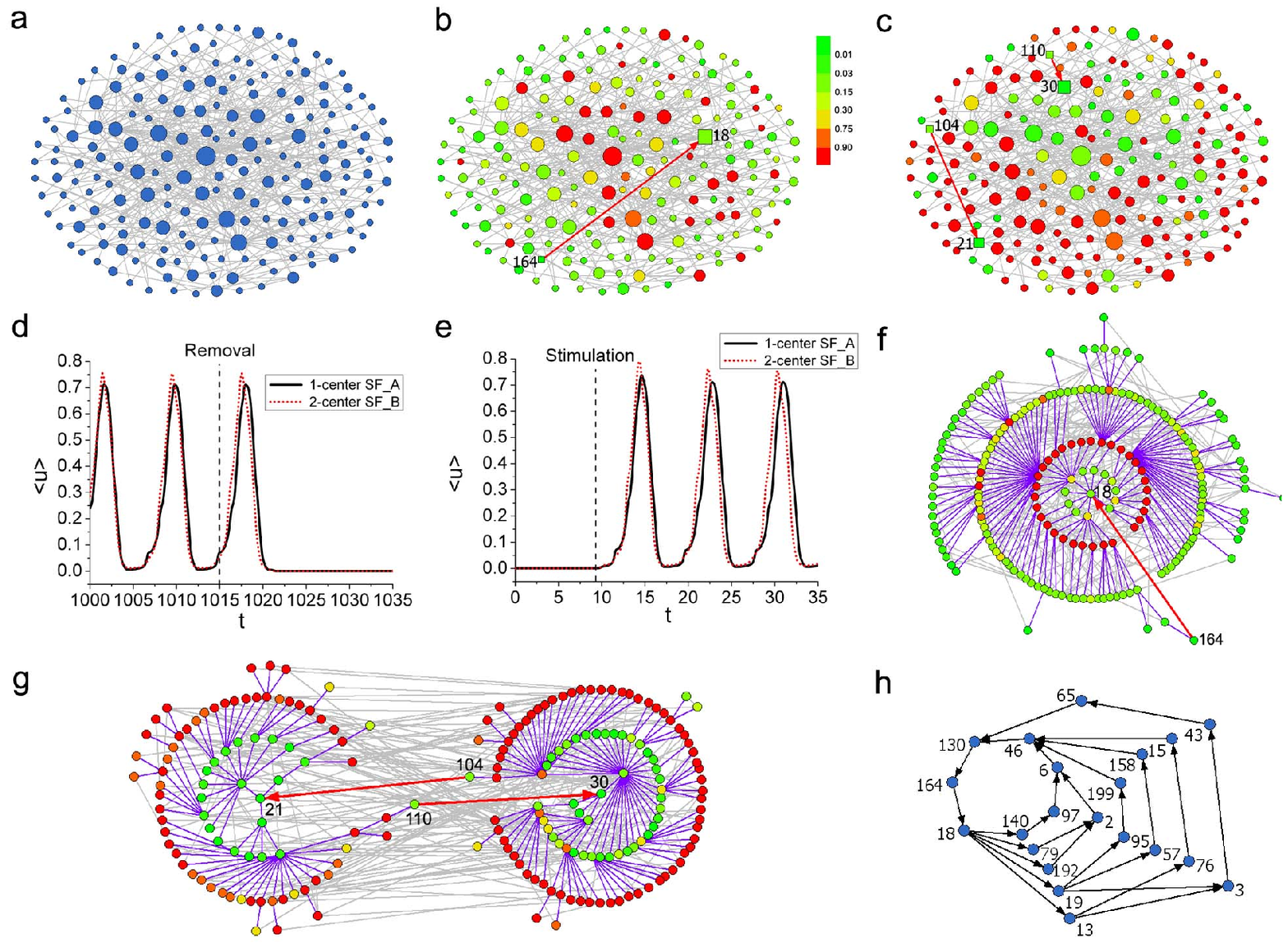}

\caption{\label{Fig6} (Color online) (a) SF network with $N=200$, $<k>=4$.
The degree distribution obeys a power-law distribution with an exponent $\gamma=-3$.
The size of each node $i$ is proportional to $ln(k_i)$. Parameters are set
as follows, $a=0.84$, $b=0.07$, $\varepsilon =0.04$, $D_u=1.0$,
$K=1.8$. (b) Snapshot of oscillation $SF_A$ with a single pair of key
nodes denoted by squares. (c) Snapshot of oscillation $SF_B$ with
two pairs of key nodes identified. (d) Suppression of oscillations
$SF_A$ and $SF_B$ by removing their center nodes ($18$ for $SF_A$
and nodes ($21$, $30$)) for $SF_B$). (e) Creation of oscillations
$SF_A$ and $SF_B$ by initially stimulating the center
nodes. (f) Snapshot same as (b) for oscillation $SF_A$, with nodes
placed in order. One-center target waves are displayed. (g) Snapshot
 same as (c) for oscillation $SF_B$ with nodes placed in order.
Because two center nodes $21$ and $30$ are almost synchronous, nodes
are arranged in the same way as Fig.\ 2(c). Two-center target waves
are observed. (h) Skeleton of oscillation $SF_A$, with all the shortest
loops with length $L=7$ displayed.}
\end{figure*}

The only difference is that due to the highly heterogeneity, there
are many short loops passing through the pair of key nodes. In the skeleton
shown in Fig.\ 6(h), only the shortest loops with $L=7$ are
demonstrated. Destroying any of the shortest loop will not
significantly change the period of the oscillation, for the remaining
shortest loops still have a length $L=7$.

So far our investigation has been performed in networks with B\"{a}r
model as local dynamics. Actually, the principles can be also
applied to other excitable systems. Here we study
Fitzhugh-Nagumo (FHN) model\cite{Fitzhugh1961} which has been used for describing the
dynamics of neural cells. Complex networks of FHN nodes with diffusive
couplings are described as follows,

$$\frac{du_i}{dt}=\frac{1}{\varepsilon}
(u_i-\frac{u_i^3}{3}-v_i)+D_u \sum_{j=1}^{N}M_{ij}(u_j-u_i)\ ,$$
$$\frac{dv_i}{dt}=\varepsilon (u_i+\beta-\gamma v_i),\quad \qquad i=1,2,...,N. \quad \eqno (2)$$

In Fig.\ 7(a) we show a homogeneous random network under investigation with $N=200$,
$k=3$. In Figs.\ 7(b)-7(h) we do the same as
Figs.\ 6(b)-6(h), respectively, with model Eq.\ (2) and network
Fig.\ 7(a) considered. Apart from the skeleton (Fig.\ 7(h)) of the one-center oscillation, the skeleton of the two-center oscillation is also demonstrated in Fig.\ 7(i). Two clusters of loops
are displayed. We find that all conclusions derived from Figs.\ 2-6 are also applicable to Fig.\ 7,
though the local dynamics and the coupling form are considerably different
from those in Eq.\ (1). Moreover, the conclusions do not depend on
the specific parameters given in Eqs.\ (1) and (2). When connective
nodes are excited in sequence, Principles (i) and (ii) are
applicable.

\begin{figure*}
\centering
\includegraphics[scale=1.0]{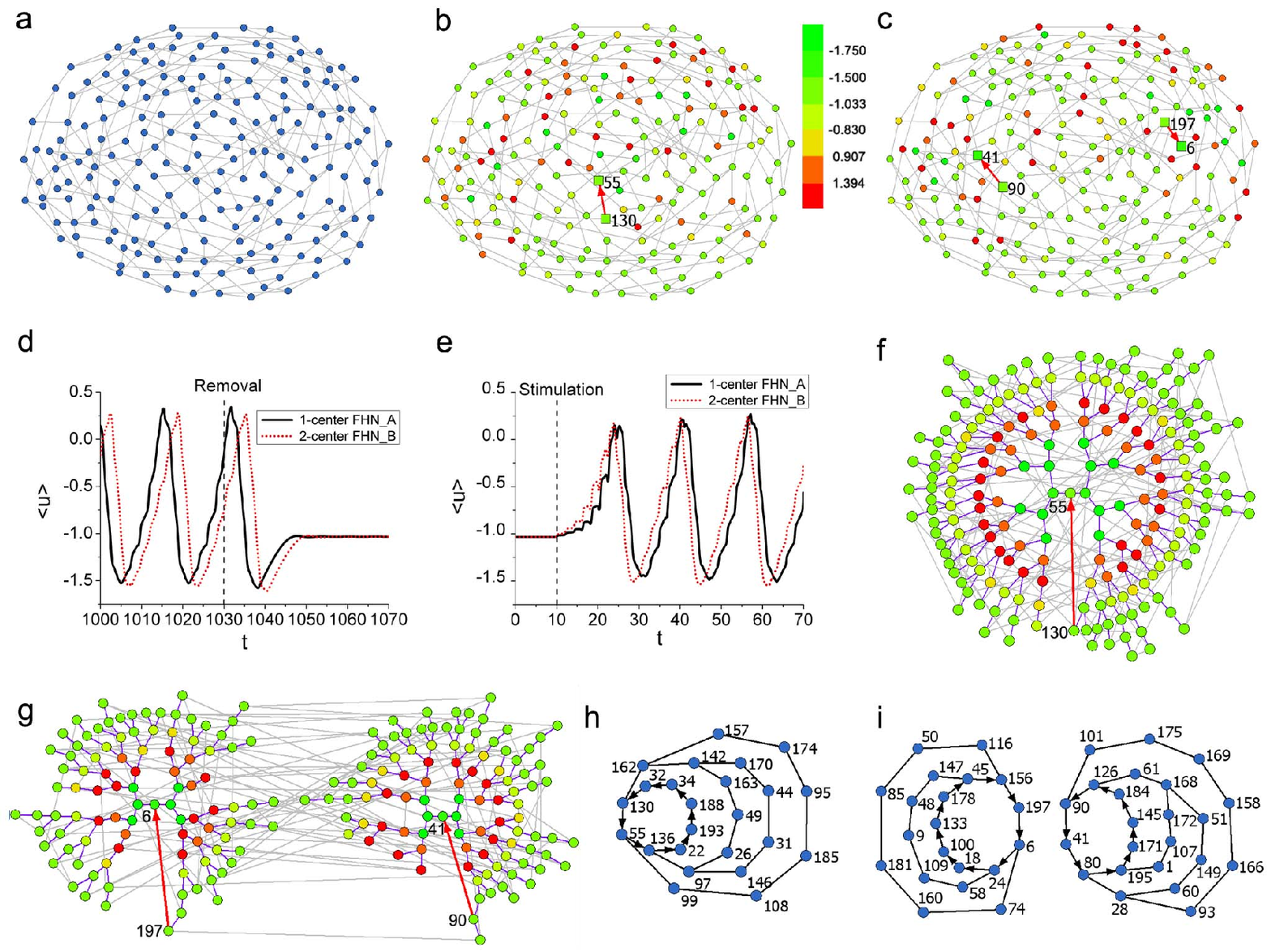}
\caption{\label{Fig7}  (Color online) (a) Homogeneous network considered
with $N=200$, $k=3$. Network dynamics is described by Eq.\ (2).
Parameters are set as follows, $\gamma=0.5$, $\beta=0.7$,
$\varepsilon=0.2$ and $D_u=0.1$. (b) Snapshot of oscillation $FHN_A$
with a single pair of key nodes. (c) Snapshot of oscillation $FHN_B$
with two pairs of key nodes. (d) Suppression of oscillations $FHN_A$
and $FHN_B$ by removing center nodes ($55$ for $FHN_A$ and nodes
($6$, $41$) for $FHN_B$). (e) Creation of oscillations $FHN_A$ and
$FHN_B$ by initially stimulating the centers nodes.
(f) Snapshot same as (b) for oscillation $FHN_A$, with nodes placed
in order. One-center target waves are displayed. (g) Snapshot same as
(c) for oscillation $FHN_B$ with nodes rearranged as Fig.\ 2(c). Two-center target waves are
observed. (h) Skeleton of oscillation $FHN_A$ with loops passing through the
pair of key nodes ($55$, $130$) with length $L \leqslant 9$. (i) Skeleton of
oscillation $FHN_B$. Two clusters of loops with $L\leqslant 10$ are displayed.}
\end{figure*}

\section{Conclusions}
In this paper we have studied pattern formation in oscillatory
complex networks consisting of excitable nodes. Well-organized
structures, including center nodes and skeletons, are revealed for seemingly random patterns.
Two simple principles are proposed: well-behaved target waves are demonstrated propagating from center nodes along the shortest paths;
the short loops passing through both the center nodes and their
drivers dominate the network oscillations. The existence of target waves with
certain centers in random networks may provide prospective insights into pattern formation in
complex networks. Moreover, the discovery of skeletons will improve
the understanding of crucial topological effects on the network dynamics.
Based on the mechanism revealed, we are able to suppress, create and modulate the
oscillatory patterns by manipulating a few nodes. All the modulations can be
predicted by analyzing the skeletons. Our surprising and useful findings are applicable to
homogeneous random networks with different sizes and degrees, inhomogeneous networks
and networks with different excitable models, such as FHN model.

In the present paper we considered periodic self-sustained oscillations
in excitable complex networks. The extensions to
nonperiodic and even chaotic oscillations will be our future work.
The ideas and methods in the present work are expected to be
applicable to wild fields where oscillatory behavior of excitable
complex networks is involved, especially for neural systems. Though
at present we do not consider some specific processes of neural
systems, we do hope that our results may have useful impact on the
investigation of complicated neural functions, since oscillatory
behavior, excitable dynamics and complexity of interactions are
crucially important for the functions of neural systems.

\begin{acknowledgments}
This work was supported by the National Natural Science Foundation
of China under Grant No.\ 10975015 ,the National Basic Research
Program of China (973 Program) (2007CB814800) and the Science Foundation of Baoji
University of Arts and Sciences under Grant No.\ ZK1048.
\end{acknowledgments}

\end{document}